\documentclass{jfm} 
\usepackage{amsmath}

\usepackage{graphicx}
\usepackage{natbib}

\ifCUPmtlplainloaded \else
  \checkfont{eurm10}
  \iffontfound
    \IfFileExists{upmath.sty}
      {\typeout{^^JFound AMS Euler Roman fonts on the system,
                   using the 'upmath' package.^^J}%
       \usepackage{upmath}}
      {\typeout{^^JFound AMS Euler Roman fonts on the system, but you
                   dont seem to have the}%
       \typeout{'upmath' package installed. JFM.cls can take advantage
                 of these fonts,^^Jif you use 'upmath' package.^^J}%
      }
  \else
  \fi
\fi

\ifCUPmtlplainloaded \else
  \checkfont{msam10}
  \iffontfound
    \IfFileExists{amssymb.sty}
      {\typeout{^^JFound AMS Symbol fonts on the system, using the
                'amssymb' package.^^J}%
       \usepackage{amssymb}%
         
         \let\geq=\geqslant
      }{}
  \fi
\fi

\ifCUPmtlplainloaded \else
  \IfFileExists{amsbsy.sty}
    {\typeout{^^JFound the 'amsbsy' package on the system, using it.^^J}%
     \usepackage{amsbsy}}
    {\providecommand\boldsymbol[1]{\text{\boldmath $##1$}}}
\fi

\newsavebox{\astrutbox}
\sbox{\astrutbox}{\rule[-5pt]{0pt}{20pt}}

\title[]{Roll convection of binary fluid mixtures in porous media}
\author[R. Umla, M. Augustin, B. Huke and M. L\"ucke]{R. \ns U\ls M\ls L\ls A, \ns%
M.\ns A\ls U\ls G\ls U\ls S\ls T\ls I\ls N,\ns 
B.\ns H\ls U\ls K\ls E\ls \break \and 
M.\ns L\ls\"U\ls C\ls K\ls E}
\affiliation{Institut f\"ur Theoretische Physik, Universit\"at des Saarlandes, Postfach 151150, D-66041 Saarbr\"ucken, Germany}
\date{\today}
\pubyear{2010}
\volume{??}
\pagerange{??}
\date{?? and in revised form ??}

\begin{document}

\maketitle

\begin{abstract}

We investigate theoretically the nonlinear state of ideal straight rolls in the Rayleigh--B\'enard system of a fluid layer heated from below with a porous medium using a Galerkin method. 
Applying the Oberbeck-Boussinesq approximation, binary mixtures with positive separation ratio are studied and compared to one-component fluids. Our results for the structural properties of roll convection resemble qualitatively the situation in the Rayleigh--B\'enard system without porous medium except for the fact that the streamlines of binary mixtures are deformed in the so-called Soret regime. The deformation of the streamlines is explained by means of the Darcy equation which is used to describe the transport of momentum. In addition to the properties of the rolls, their stability against arbitrary infinitesimal perturbations is investigated. We compute stability balloons for the pure fluid case as well as for a wide parameter range of Lewis numbers and separation ratios which are typical for binary gas and fluid mixtures. The stability regions of rolls are found to be restricted by a crossroll, a zigzag and a new type of oscillatory instability mechanism, which can be related to the crossroll mechanism. 
\end{abstract}

\section{Introduction}
The Rayleigh--B\'enard system is one of the classical setups and possibly the most popular one to study pattern formation and hydrodynamic instabilities under well-controlled conditions. The system consists of a fluid layer bounded by two plates and is heated from below. If the temperature difference between the plates exceeds a critical value, the fluid leaves its static state and thermal convection sets in. Under appropriate conditions, the fluid flow forms a regular pattern, for example a state of ideal straight rolls (ISR), also called B\'enard rolls or steady overtuning convection (SOC).
 
Over the last decades, the roll convection and in particular the instability mechanisms, which tend to limit the stability region of the ideal roll state, have been investigated in detail experimentally (see \cite{Busse71} and \cite{Croquette89a}) as well as theoretically (see \cite{BusseClever,Schlueter65,ClBu90,Bo86,Bolton85}). One main result of this research is that the region of rolls, which are stable against infinitesimal perturbations, is restricted by certain instability mechanisms, for instance the zigzag, the skewed varicose, the Eckhaus and the crossroll mechanism. This remains also valid if the fluid between both the plates saturates a porous medium, a case applying to many natural and industrial processes (see \cite{Vaf05} or \cite{Ni06} for an overview). For a porous medium,  \cite{Torre} investigated theoretically the stability of rolls against two-dimensional perturbations and found the Eckhaus and an oscillatory instability mechanism. \cite{Straus} did the same but allowed for three-dimensional perturbations and found a zigzag as well as a crossroll mechanism. These investigations were complemented by the experimental observation of roll convection in porous media by \cite{Sha97} and \cite{How97}.
 
The aforementioned research was carried out for a one-component, i.e., a pure fluid. However, numerous applications are known where the fluid has to be treated as a binary mixture consisting of two different fluids \cite[see for example][]{Pl06}. Especially, if the binary mixture shows a non-vanishing Soret effect, i.e., if concentration currents are driven by temperature gradients, the dynamics of the system are expected to change due to a coupling of the temperature field into the concentration field. For the classical Rayleigh--B\'enard system without porous medium, there has been a large body of work dealing with binary mixtures under the influence of the Soret effect (\cite{Eaton,Barten1,Huke2,Gal,Lerma1,Barten2,Schoepf,KnoblochMoore,CrossKim,KolodnerPRL,Fuetterer1,Touiri,Ahlers2,Walden}). In particular, the roll convection and the corresponding stability mechanisms are well known (see \cite{Huke1}). For the system with porous medium, however, the standard of knowledge is much less developed, although there has been some work concerning the stability of the ground state and monocellular flow by \cite{Mo07}, \cite{Mo08} and \cite{So01}.
  
With this paper, we aim at extending the knowledge about flow patterns of binary mixtures in porous media by analysing the structural properties as well as the stability of the B\'enard rolls. In doing this, we restrict ourselves to a positive Soret coupling in which the Soret effect destabilises the ground state. The paper is organised as follows: In Sec. II we will briefly explain the basic equations of the system, the Galerkin method that we have used for numerical computations, the ground state and its stability. In Sec. III we discuss the properties of roll convection by investigating the structure of the fields, the advective heat transport and the mixing. Sec. IV contains the main results of our paper. Here,  we represent the stability boundaries of the B\'enard rolls for a pure fluid as well as for binary mixtures and compare the former ones to the known results. In Sec. V we conclude with a summary of our results.

\section{Foundations}
\subsection{System and basic equations}
We consider a horizontal layer of a porous medium filled with a pure fluid or a binary mixture in a homogeneous gravitational field, $ \textbf{g}=-g\textbf{e}_{z}$. The layer has thickness $d$ and a vertical temperature gradient is imposed by fixing the temperature 
\begin{eqnarray} T = T_{0}\pm\frac{\Delta T}{2} \; \; \; \text{at} \;\;\; z=\mp\frac{d}{2},\end{eqnarray}
which can be realized in experiments by using highly conducting plates. $T_{0}$ is the mean temperature of the layer and we assume $\Delta T>0$, i.e., the lower plate has a higher temperature than the upper one. Moreover, we consider the plates to be infinitely extended, rigid and impermeable. The porous medium is treated as isotropic and homogeneous. Furthermore, we assume local thermal equilibrium between the fluid and the porous medium such that there is no heat transfer between both phases.
 
Convection is characterised by the fields of temperature $T$, the Darcy velocity $\textbf{v}$ (also called seepage velocity), mass concentration $C$ of the lighter component, total mass density $\rho=\rho_{1}+\rho_{2}$ and pressure $P$. Applying the Oberbeck-Boussinesq approximation, we assume the dynamic viscosity, the thermal expansion coefficients and the heat capacities to be constant and equal to their values at the spatial averages $T_{0}$, $C_{0}$, $P_{0}$ of the thermodynamic variables. Moreover we neglect effects like radiation or the generating of heat due to friction between the fluid and the porous medium. For binary mixtures, the Soret effect can play an important role. We incorporate this effect in our model, which describes the generation of concentration currents due to temperature variations. Then, the balance equations for our system read according to \cite{Ni06}:
\begin{subequations}
\begin{eqnarray}
  \nabla \cdot \textbf{v} & = & 0 \\ 
  c_a \rho_0 \partial_t {\bf v } & = & - \nabla P - \frac{\eta}{K}\textbf{v} + \rho_{0}[1+\beta_T T+\beta_C C]g{\bf e}_z\\
C_{\text{tot}} \partial_t T+ C_f ( {\bf v} \cdot \nabla) T & = &\lambda_{\text{tot}} \nabla^2 T \\
\phi \partial_t C + ({\bf v} \cdot \nabla) C & = & D_{\text{tot}} \nabla^{2} C + D_{\text{tot}} \frac{k_{T}}{T_0}  \nabla^{2} T, 
\end{eqnarray}
\label{basiceq}
\end{subequations}
Here, $\rho_{0}$ is the mean density of the fluid, $\eta$ the dynamic viscosity, $\beta_{T}$ ($\beta_{C}$) the thermal (solutal) expansion coefficient of the fluid and $K$ the permeability of the porous medium. $C_*$ denotes the heat capacity per unit volume according to its subscript $f$, $s$ or $\text{tot}$ of the fluid, the solid matrix or of the total medium (fluid and solid matrix). The heat capacities per unit volume can be connected via the porosity $\phi$ of the porous medium, by the relation $C_{\text{tot}}=\phi C_{f} + (1-\phi)C_{s}$. The same holds true for the thermal conductivities $\lambda_{f}$, $\lambda_{s}$ and $\lambda_{\text{tot}}$. $D_{\text{tot}}$ is the concentration diffusivity of the total medium and is equal to $\phi D$ whereas $D$ is the concentration diffusivity of the fluid. $k_{T}$ is the thermodiffusion ratio which characterises the strength of the Soret effect. The correction factor $c_a$ emerges in front of $\partial_t \textbf{u}$ to bring the Darcy equation $(\ref{basiceq}b)$ in agreement with experimental results (see \cite{Ni06}). Since the porous medium is assumed to be isotropic and homogeneous, $c_a$ is a scalar. 
 
In the basic equations (\ref{basiceq}) we scale lengths by $d$, time by the vertical diffusion time $\frac{d^{2}C_{\text{tot}}}{\lambda_{\text{tot}}}$, temperature by $\frac{\lambda_{\text{tot}}\eta}{C_{f}Kd\beta_{T}g\rho_{0}}$,  concentration by $\frac{ \lambda_{\text{tot}}\eta}{C_{f}Kd\beta_{C}g\rho_{0} }$ and pressure by $\frac{\lambda_{\text{tot}}\eta}{ C_{f}K}$. Moreover, we introduce the reduced deviations of the velocity $\textbf{u}=(u,v,w)$,  temperature $\theta$, concentration $c$ and pressure field $p$ from the conductive state thus obtaining the following set of balance equations:
\begin{subequations}
\begin{eqnarray}
  \nabla \cdot \textbf{u} & = & 0 \\ 
  \gamma_a \partial_t {\bf u } & = & - \nabla p - \textbf{u} + ( \theta + c ) {\bf e}_z\\
\partial_t \theta + ( {\bf u} \cdot \nabla) \theta & = & R w + \nabla^2 \theta \\
{\phi^*} \partial_t c + ({\bf u} \cdot \nabla) c & = & R \psi w + L \left( \nabla^2 c - \psi \nabla^2 \theta \right). 
\end{eqnarray}
\label{bal}
\end{subequations}
The normalised porosity $\phi^{*}=\phi C_{f}/C_{\text{tot}}$ as well as the correction factor $\gamma_{a} = \frac{K \lambda_{\text{tot}} \eta}{d^2 C_{\text{tot}}\rho_{0}} c_{a}$ are scalars for an isotropic and homogeneous porous medium. $\phi^{*}$ can have values in the interval $[0,1]$ whereas $\gamma_{a}$ is very small for usual porous materials (\cite{Ni06}). Because of the smallness of  $\gamma_{a}$, the time-derivative term in the momentum equation is often neglected (see for example  \cite{So01} and \cite{Mo08}). However, we retain this term since it is - according to \cite{Vad99} or \cite{Vad00} - essential for certain stability analyses. Note that $\phi^{*}$ also appears only in front of a time derivative. Therefore, the correction factor and the normalised porosity can play a role for time-dependent phenomena only. 

Via the concentration diffusivity $D_{\text{tot}}$, the Lewis number $L$ is defined by 
\begin{eqnarray}L=\frac{D_{\text{tot}}\phi C_{f}}{\lambda_{\text{tot}}}, \end{eqnarray}
and therefore compares the time scales of concentration and heat diffusion. Note that the Lewis number as defined here is the inverse of the Lewis number $Le$ as defined by \cite{Ni06}. The Rayleigh--Darcy number is given by 
\begin{eqnarray}R=\frac{\rho_{0} g \beta_{T} K C_{f} d}{\lambda_{\text{tot}} \eta}\Delta T \end{eqnarray}
and measures the thermal driving. The separation ratio 
\begin{eqnarray}\psi=-\frac{\beta_{C}}{\beta_{T}T_{0}}k_{T} \end{eqnarray}
is proportional to the thermodiffusion ratio $k_{T}$ thus incorporating the Soret effect which emerges via $-L\psi \nabla^2 \theta$ and the term $R \psi w$ in the concentration balance equation (\ref{bal}d). In this paper, we cover mainly separation ratios from $0$ to $0.5$, which are typical for alcohol-water mixtures at high alcohol concentration (30-80\%) and many gas mixtures, like $\text{Ne-CO}_{2}$, He-Xe, $\text{H}_{2}\text{-Xe}$ and Ne-Ar (see \cite{Liu}). That means in particular that we restrict ourselves to a positive Soret coupling ($\psi>0$), where the lighter component of the mixture is driven into the direction of higher temperature.

In the momentum equation (\ref{bal}b), which is deduced from Darcy's law, advective transport and diffusion of momentum are neglected. Instead, we take into account the relaxation term $-\textbf{u}$ describing the friction between the fluid and the porous matrix, which dominates the aforementioned mechanisms at low Reynolds numbers ($Re=O(1)$). Thereby, to define the Reynolds number of a flow through a porous medium, the average radius of the pores is chosen as the characteristic length scale. 

Since the fluid is assumed to be incompressible, we can write the velocity field $\textbf{u}$ as
\begin{eqnarray} \textbf{u} = \nabla \times \nabla \times \Phi \textbf{e}_{z}+ \nabla \times \Psi \textbf{e}_{z},
\end{eqnarray}
which automatically fulfils the mass balance (\ref{bal}a). To derive equations for the potentials $\Phi$ and $\Psi$, one can apply the curl to the momentum balance equation (\ref{bal}b) once or twice, respectively and take into account only the third component of the obtained equations. This also eliminates the pressure term. To derive boundary conditions for $\Phi$, we assume that there is no external pressure gradient which causes a meanflow $U(z,t)=\langle u(x,z,t) \rangle_{x}$, with $\langle\cdot\rangle_{x}$ being the spatial average in $x$--direction. Averaging (\ref{basiceq}b) over $x$ shows that any meanflow will decay on the time scale $\gamma_{a}$. The same can be demonstrated for $\Psi$. Not being interested in fast transients, we assume that $U=\Psi=0$.  \newline
At the impermeable plates, the vertical concentration current vanishes, i.e., $\partial_{z}(c-\psi \theta)=0$ for $\pm\frac{1}{2}$. To avoid this coupling of the fields in the boundary conditions, we introduce in a next step the field
\begin{equation} \zeta=c-\psi\theta. \end{equation}
 The corresponding balance equation for $\zeta$ can directly be derived by combining the equations (\ref{bal}c) and (\ref{bal}d).
Altogether, the new balance equations read 
\begin{subequations}
\begin{eqnarray}
\gamma_{a} \partial_{t}\nabla^{2}\Delta_{xy}\Phi & = &-\nabla^{2}\Delta_{xy}\Phi -\Delta_{xy} ( \theta + c ) \\
\partial_t \theta + ( {\bf u} \cdot \nabla) \theta & = &  R w + \nabla^2 \theta \\
\!\!\!\!\!\!\!\!\!\!\!\!\!\!\!\!\ \phi^{*} \partial_{t} \zeta + \left(\mathbf{u} \cdot \mathbf{\nabla}\right) \zeta + \left(1 - \phi^{*}\right) \left(\mathbf{u} \cdot \mathbf{\nabla}\right) \theta & = &  \left( \phi^{*}- 1\right) R \psi \Delta_{xy} \Phi + L \mathbf{\nabla}^{2} \zeta - \phi^{*} \psi \mathbf{\nabla}^{2} \theta 
\end{eqnarray} 
\label{balnew}
\end{subequations}
with $\Delta_{xy}:= \partial_{x}^{2}+\partial_{y}^{2}$ and the boundary conditions
\begin{eqnarray}0=\Phi=\theta=\partial_{z}\zeta \; \; \;  \text{at} \; \; z=\pm\frac{1}{2}.\label{bdcond} \end{eqnarray}

\subsection{Numerical method}
\label{num}
To obtain roll solutions, we used a Galerkin method with the following ansatz for the fields $X=\Phi,\theta,\zeta$: 
\begin{eqnarray}X(x,z)=\sum_{m,n}X_{mn}\text{cos}(mkx)f_{n}(z). \label{ansatz-roll}\end{eqnarray} 
This ansatz is almost the same as in \cite{Huke1} where it is described in more detail.\newline
 The difference is that $\psi\equiv 0$ and that $\Phi$ is expanded in the vertical direction by 
\begin{eqnarray}
 \Phi: f_{n}(z)=\begin{cases}\sqrt{2}\;\text{cos}(n\pi z) &\text{if} \;\; n=1,3,5...\\ 
                  					           \sqrt{2}\;\text{sin}(n\pi z) &\text{if} \;\; n=2,4,6...
					              \end{cases}
\label{vert1}
\end{eqnarray}
to satisfy the boundary condition (\ref{bdcond}). 
Rolls are even in $x$ with an appropriate choice of the origin and fulfil the so-called mirror-glide symmetry (see \cite{Ve66}). This allows to simplify the ansatz (\ref{ansatz-roll}) by dropping half the modes. In the following computations,  all $\theta$- and $\zeta$-modes with $n+m>N$ and all $\Phi$-modes with $n+m>N/2$ are neglected, where we choose $N=24$ if not otherwise specified.\newline
The stability of rolls against arbitrary infinitesimal perturbations is tested by expanding the perturbations as follows
 \begin{align}\delta X(x,y,z,t)=\!\sum_{m,n}\delta X_{mn}e^{st+i[(d-mk)x+by]}f_{n}(z),\end{align}
where the ansatz is truncated in a way consistent with the truncation of (\ref{ansatz-roll}). 
Except for the differences pointed out already, this ansatz and the corresponding method can be found in \cite{Huke1}.  The symmetries of the underlying roll pattern allow for $b=0$ to divide the perturbations into perturbations that are even or odd in the x-direction and into $G$-- as well as $\overline{G}$-perturbations, that reduce to symmetric and antisymmetric perturbations under the mirror glide operation in the special case d=b=0.
Moreover,  because of these symmetries, only perturbations with $b>0$ and $d\in[0,k/2)$ have to be tested.

\subsection{Ground state and linear Stability Analysis}
\label{Ground}
The linear stability problem for binary mixtures of the system under consideration has already been investigated by \cite{So01}, \cite{Mo07} and \cite{Mo08}. Having used an analytical method and a Galerkin expansion, we obtain results in good agreement with the former. In the following, we are going to review the most important facts concerning the conductive state as a preparation and introduction for the subsequent nonlinear analysis.

In the ground state, the fluid rests and heat is transported only by diffusion. The temperature difference $\Delta T$ between the plates imposes a linear temperature profile as follows: 
 \begin{eqnarray}T_{\text{cond}}(z)=T_{0}-\frac{\Delta T}{d}z.\end{eqnarray}
Due to the Soret effect, this temperature gradient generates a linear concentration profile of the form: 
 \begin{eqnarray}C_{\text{cond}}(z)=C_{0}+\frac{k_{T}\Delta T}{T_{0}d}z.\end{eqnarray}
The pressure distribution in the conductive state is given by 
\begin{eqnarray}p_{\text{cond}}(z)=p_{0}+\rho g\left[ \beta_{T}(T_{0}-\frac{\Delta T}{2d}z)+ \beta_{C}(C_{0}-\frac{k_{T}\Delta T}{2dT_{0}}z)\right]z,\end{eqnarray}
which can be calculated from the concentration and temperature field via the momentum equation (\ref{bal}b).
If there is a positive Soret coupling as investigated here, the lighter component of the binary fluid is driven to the lower plate. This increases the density difference between the plates and therefore destabilises the ground state.
 
\begin{figure}
\includegraphics[width=0.95\textwidth]{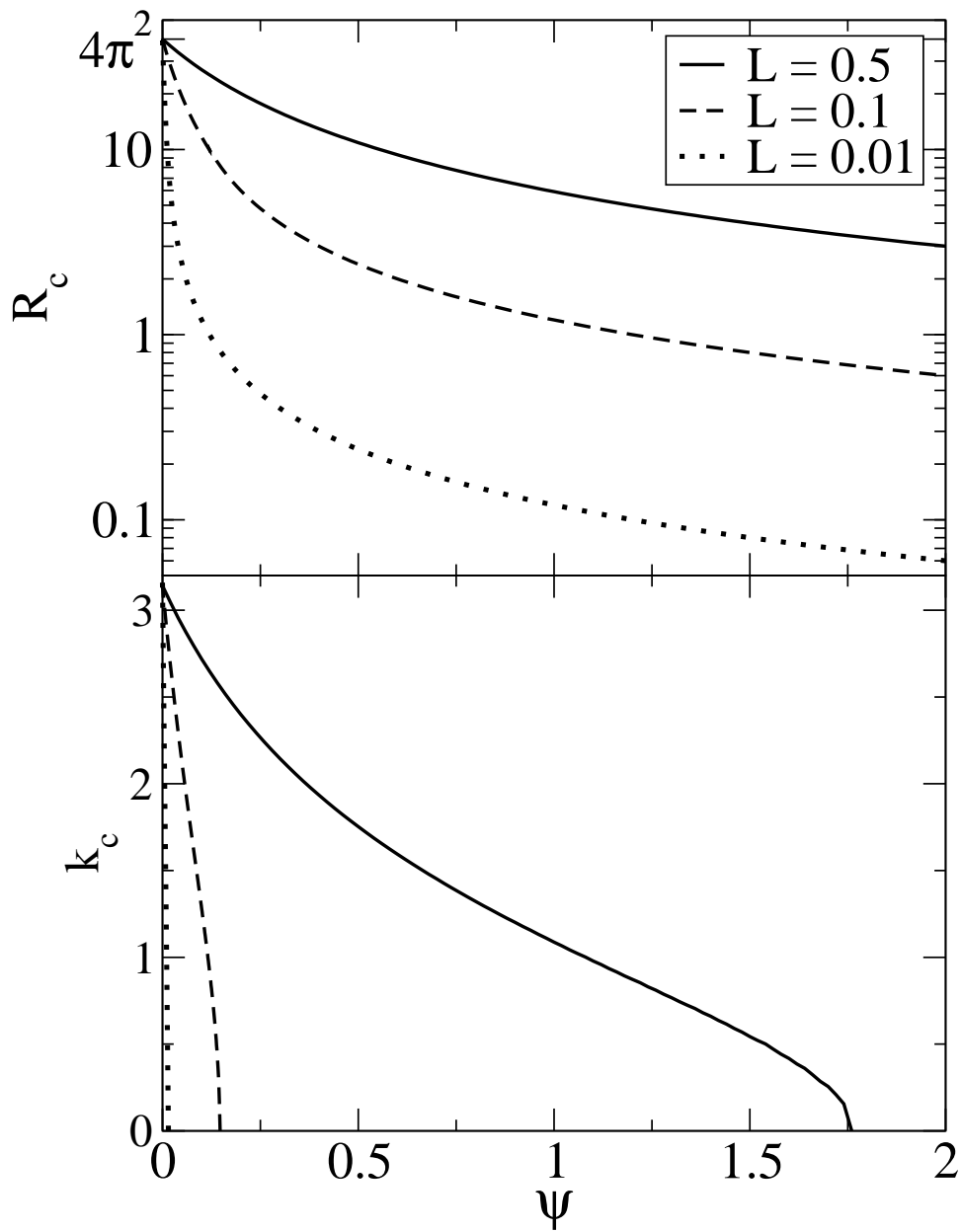}
\caption{In the upper (lower) part, the critical Rayleigh--Darcy number of the ground state (corresponding critical wavenumber) is plotted against the separation ratio $\psi$ for several $L$. If the thermal driving is weaker than the critical value, the conductive state is stable against arbitrary infinitesimal perturbations, otherwise at least one unstable mode exist. }
\label{G} 
\end{figure}

The destabilising effect is illustrated in Figure \ref{G} where the critical Rayleigh--Darcy number $R_{c}$ as well as the corresponding critical wavenumber $k_{c}$ is  plotted against the separation ratio $\psi$ for several $L$. For the pure fluid ($\psi=0$), the ground state loses its stability above $R_{c}^{0}=4\pi^{2}\approx 39.48$ against stationary perturbations of a lateral wavenumber $k_{c}^{0}=\pi$. By contrast, the ground state of binary mixtures becomes already unstable for a thermal driving weaker than $R_{c}^{0}$ and the critical Rayleigh--Darcy number decreases with stronger Soret effect, i.e., growing $\psi$. In particular, if the concentration gradients are slowly diffused away, i.e., for small Lewis numbers, the destabilisation is especially strong. The corresponding critical perturbations remain stationary but their wavelength $k_{c}$ goes to zero for large $\psi$. For example at $L=0.5$, one finds $k_{c}=0$ for about $\psi\geq1.759$ which is close to the theoretical value $\psi \geq \frac{1}{\frac{40}{51 L}-1} = 1.7586$ obtained by \cite{So01}. In experiments, the critical wavelength will then be as large as the finite size of the convection cell allows.

\section{Properties of the roll convection}
\subsection{Structure of the fields}
\begin{figure}
\includegraphics[width=0.95\textwidth]{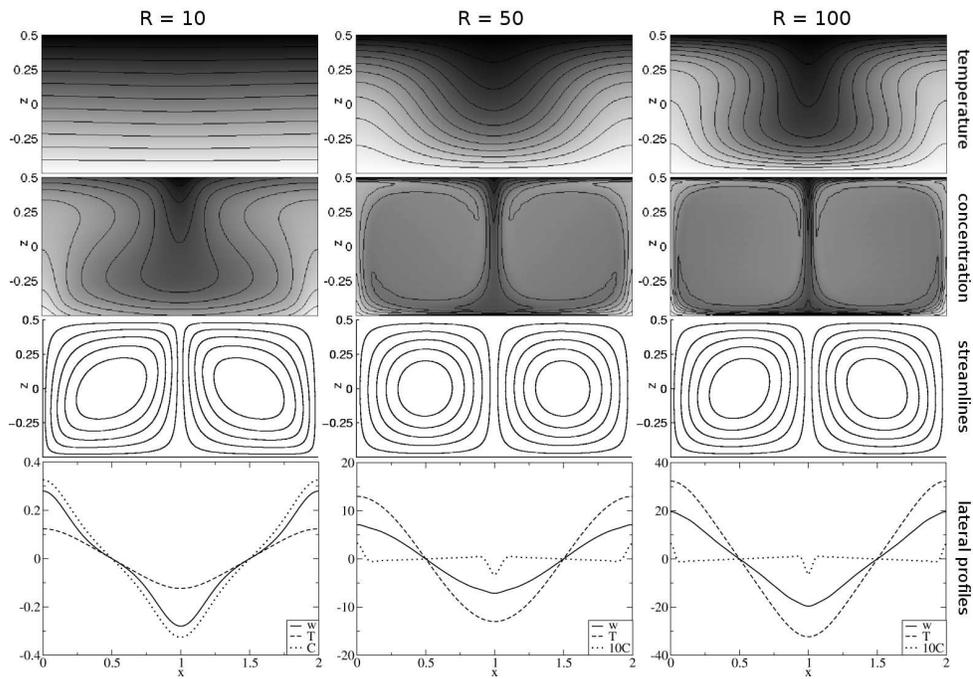}
\caption{Structure of roll convection. Shown are from top to bottom the fields of temperature and concentration, the streamlines as well as the lateral profiles at midheight, $z=0$. Parameters are $\psi=0.3$, $L=0.01$, $k=\pi$. }
\label{fields} 
\end{figure}

To understand roll convection of a binary mixture from a qualitative point of view, one can study the temperature and concentration distribution as well as the flow field. In what follows, we give such a qualitative description by Figure \ref{fields}, in which the streamlines, the concentration and  temperature field along with the lateral profiles of the fields at midheight $z=0$ are shown for several values of thermal driving. Since the behaviour remains in principle the same over a wide range of wavenumbers, we fix the wavenumber to $k=k_{c}^{0}=\pi$. The chosen parameters, $L=0.01$, $\psi=0.3$, can be realized easily in experiments with alcohol-water mixtures. In this case, the critical Rayleigh--Darcy number $R_{c}$ is reduced to about $0.39$. The left column in Figure \ref{fields} displays the fields over one periodicity interval in $x$-direction for $R=10$. This value lies in the so-called Soret regime where convection is dominated by the Soret effect. The temperature field deviates only marginally from its linear profile in the ground state, i.e., advective heat transport is weak. In contrast,  the concentration field is already strongly modulated and forms plume-like structures since the slow concentration diffusion ($L\ll 1$) allows for a perturbation of the linear profile already for weak advection. The anharmonicity of the concentration field becomes obvious by looking at the lower part of Figure \ref{fields}, where the lateral profile of the concentration field is plotted versus $x$ at midheight, $z=0$. The velocity field, represented by its $z$-component at midheight, is also anharmonic and the streamlines that illustrate the roll-like flow are deformed. 

The middle column of Figure \ref{fields} refers to $R=50$. This value lies in the so-called Rayleigh regime where convection would set in also without an operating Soret effect. Here, a stronger modulation of the temperature field can be seen due to an increased advective heat transport. Nevertheless, the temperature field is almost harmonic as seen from the horizontal variation of $T$ at midheight. The concentration field shows the characteristic boundary layer behaviour: The binary fluid is well mixed in the bulk whereas pronounced concentration gradients exist at the plates and the roll boundaries. The better mixing is caused by the larger velocity of the fluid. The corresponding streamlines reflect the roll-like flow and are almost harmonic indicating that the anharmonic behaviour of the velocity for $R=10$ is induced by the Soret effect. This conclusion has been strengthened when we simulated the streamlines of a pure fluid and did not find any deformation provided the thermal driving was not too strong. Note furthermore that the deformation of the streamlines in the Soret regime seems to be generic for the system with porous medium. This can be explained by the replacement of the momentum diffusion term $\nabla^{2}\textbf{u}$ (contained in the Navier--Stokes equations) by the relaxation term $-\textbf{u}$ in our momentum balance equation ($\ref{bal}b$). In the clear fluid, i.e., in the system without porous medium, momentum diffusion tends to smooth the spatial anharmonicity of the  velocity field caused by the very anharmonic concentration field. Since the relaxation term $-\textbf{u}$ in ($\ref{bal}b$) does not contain spatial derivatives, it does not provide spatial smoothing and the anharmonicity of the concentration field is imposed via the buoyancy term onto the velocity field. Taking into account that the buoyancy at $R=50$ is caused rather by the almost harmonic temperature field than by the concentration field, it becomes clear why the streamlines are barely deformed.

When nonlinear effects are amplified further by increasing the heating rate, plume-like structures appear also in the temperature field and the concentration boundary layers become thinner. We can see this in the right column of Figure \ref{fields}, where $R$ has reached a value of $100$. The streamlines are now deformed again, this time because the temperature field has become anharmonic by the intensive thermal driving.

\subsection{Nusselt and mixing numbers}
To describe the advective heat transport due to roll convection quantitatively, we make use of the Nusselt number Nu. The Nusselt number is given by 
\begin{eqnarray}\text{Nu}:= \frac{\langle j_{\text{tot},z}\rangle}{\langle j_{\text{cond},z}\rangle}\end{eqnarray}
where $j_{\text{tot},z}$ denotes the total vertical heat current density and $j_{\text{cond},z}$ the vertical heat current density in the conductive state.  $\langle \cdot\rangle$ denotes the lateral average. The Nusselt number at the plates can be computed from the modes obtained by our Galerkin method with truncation index $N$ as follows: 
\begin{eqnarray} \text{Nu}= 1- \frac{2\sqrt{2}\pi}{R}\sum_{n=1}^{N}(-1)^{n}n\theta_{02n}.\end{eqnarray} 
Note that the Nusselt number does not actually depend on the $z$-position in a stationary state of convection as the rolls are. 

\begin{figure}
\includegraphics[width=0.95\textwidth]{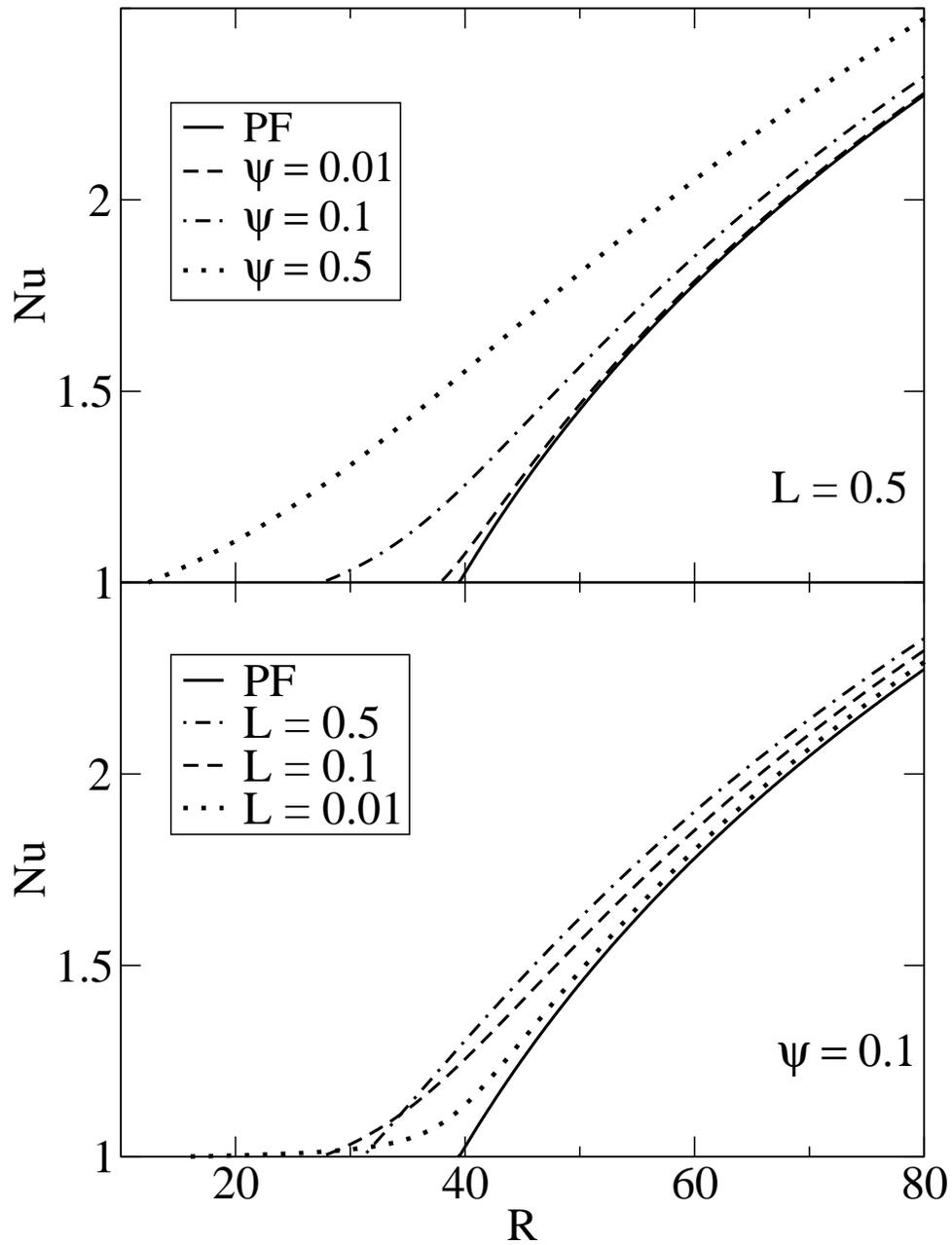}
\caption{Nusselt number Nu versus Rayleigh--Darcy number R. The parameters are $L=0.5$ in the top figure and $\psi=0.1$ in the bottom one. In each case the wavenumber is $k=\pi$.  PF denotes the pure fluid. }
\label{Nu} 
\end{figure}

In Figure \ref{Nu} we compare Nusselt numbers of the pure fluid, which we have found to be in good agreement with the results from \cite{Torre}, to Nusselt numbers of binary mixtures. In the ground state, Nu is equal to 1, since heat is transported only by diffusion. After the roll convection has started, i.e., $R$ has exceeded $R_{c}$, the Nusselt number and simultaneously the advective heat transport increases monotonically with $R$. For a given Rayleigh--Darcy number, Nu is always larger for binary mixtures with $\psi>0$ than for the pure fluid, as the Soret effect causes a concentration gradient   giving rise to a more pronounced buoyancy and thus a stronger convection. For the same reason, the Nusselt number grows when the Soret effect becomes 
As the advective mixing is strong in the Rayleigh region, the Nusselt number of a binary mixture with small $L$ approaches the one of the pure fluid there. However, when $L$ becomes larger, i.e., when concentration diffusion is fast, the advective mixing does not succeed to mix away the concentration gradient and the differences in Nu to the pure fluid case remain more significant.\newline

 We have also found these tendencies for wavenumbers that differ from the wavenumber $k=\pi$ chosen in Figure \ref{Nu}, albeit convection becomes weaker and Nu decreases if the wavenumber gets too large or too small. 
 
\begin{figure}
\includegraphics[width=0.95\textwidth]{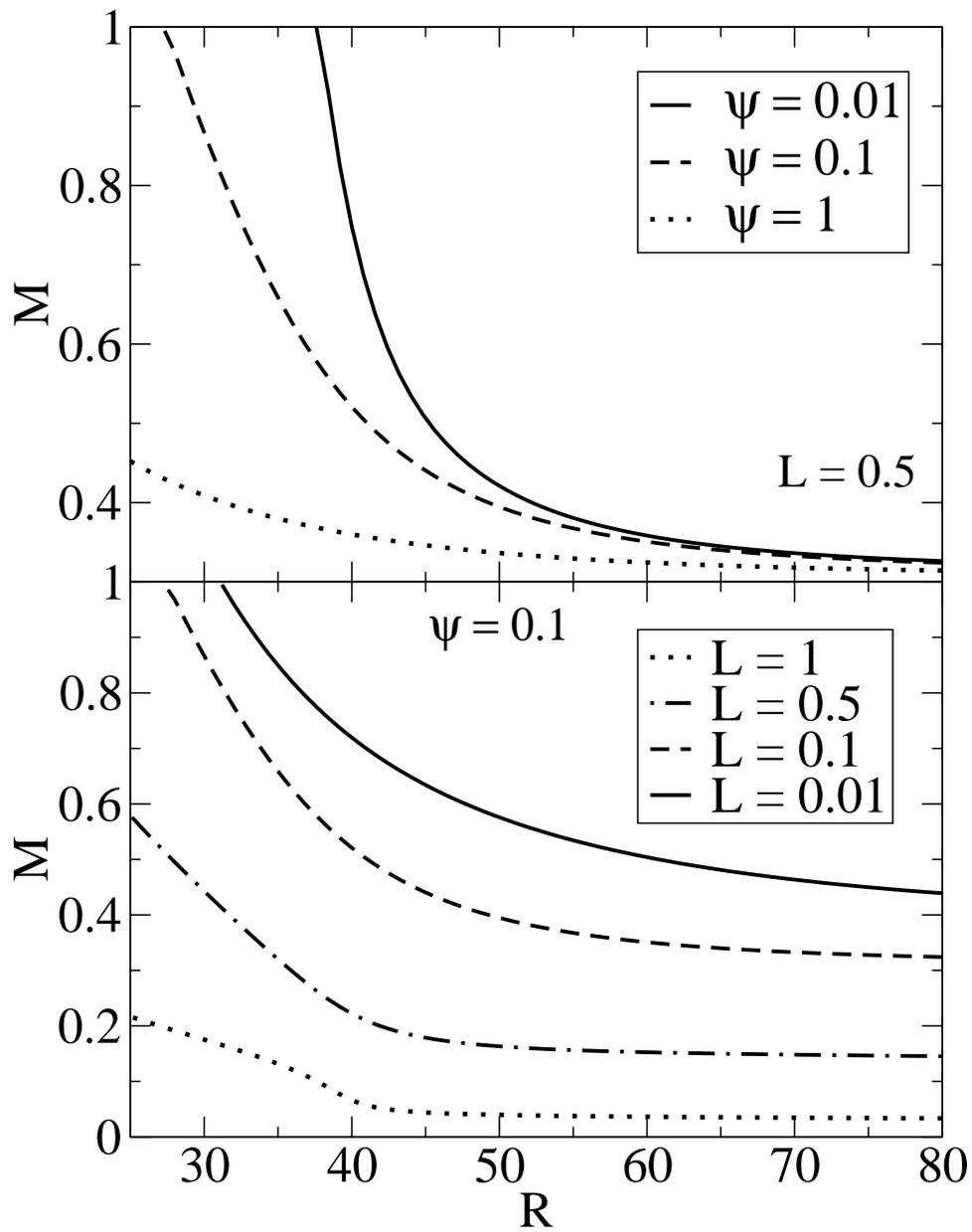}
\caption{Mixing number $M$ versus R. The parameters are $k=\pi$ and $L=0.5$ top, $\psi=0.1$ bottom.}
\label{M} 
\end{figure}

The mixing of a binary fluid can be described by the so-called mixing number $M$, which is defined by the normalised variance of the concentration field:
\begin{eqnarray}M=\frac{\sqrt{\langle C^{2} \rangle -\langle C\rangle^{2}}}{\sqrt{\langle C_{\text{cond}}^{2} \rangle - \langle C_{\text{cond}}\rangle^{2}}}.\end{eqnarray} 
$C_{\text{cond}}$ denotes the concentration field in the conductive state and $\langle\cdot\rangle$ the spatial average.
From $M=1$, which is the value of the mixing number in the ground state by definition, it decreases when roll convection sets in and the components of the binary fluid are mixed. According to Figure \ref{M}, the mixing number decreases only slightly in the Rayleigh regime  when the thermal driving gets stronger. In other words, one cannot  mix the fluid components perfectly by increasing $R$. The reason is that the fluid is already well mixed in the Rayleigh regime except for the fluid layers near to the plates and at the roll boundaries, whose thickness depends barely on $R$. Instead, the boundary layer behaviour in the Rayleigh region depends mainly on the Lewis number of the mixture such that the mixing number is reduced when $L$ is lowered. The latter facts are also valid in the clear fluid case and can be explained using concentration boundary layer theory in analogy to \cite{DissHollinger}. A larger separation ratio improves the mixing mainly in the Soret regime whereas the influence of $\psi$ in the Rayleigh regime is rather weak.

\section{Stability of the rolls}
\subsection{Stability boundaries for the pure fluid}
Using the Galerkin method from section \ref{num} with $N=24$, we have tested the roll structure against infinitesimal perturbations of arbitrary wavenumber.
For the pure fluid, similar stability analyses have been carried out by \cite{Torre} and \cite{Straus}. In agreement with their results, we find that only two instability mechanisms, the zigzag and the crossroll mechanism, limit the region of stable rolls. The corresponding stability boundaries in the $(R,k)$-parameter space are shown in Figure \ref{stab_psi=0}. Below, we review briefly the important instability mechanisms and compare our results to those of Straus and Ju\'arez. If not otherwise specified, the instabilities are related to a real eigenvalue, i.e., the corresponding stability boundary does not depend  on $\phi^{*}$ or $\gamma_{a}$.

\subsubsection{Zigzag instability }
The zigzag (ZZ) boundary, denoted by the dash-dotted line in Figure \ref{stab_psi=0}, restricts the region of stable rolls on the small-$k$ side; if the rolls are zigzag-unstable, a new set of rolls with a larger wavenumber begins to grow. The ZZ perturbations belong to the subclass of $G$-perturbations that are odd in $x$-direction. They have the same periodicity in $x$-direction as the existing pattern. Consequently, we find the ZZ instability for $d=0$. It is sufficient to check the stability only for a single point on the b-axis near $b=0$ in order to determine whether a roll state is ZZ-stable or not (for further details see \cite{Bo86}).

Compared to the ZZ boundary calculated by Straus, ours is much more restrictive. At $R=55$ for example, rolls should be unstable for $k<3.046$ according to our computations whereas Straus states the same for $k< 2.5$ using an analytical criterion.

\subsubsection{Crossroll instability}
\begin{figure}
\includegraphics[width=0.95\textwidth]{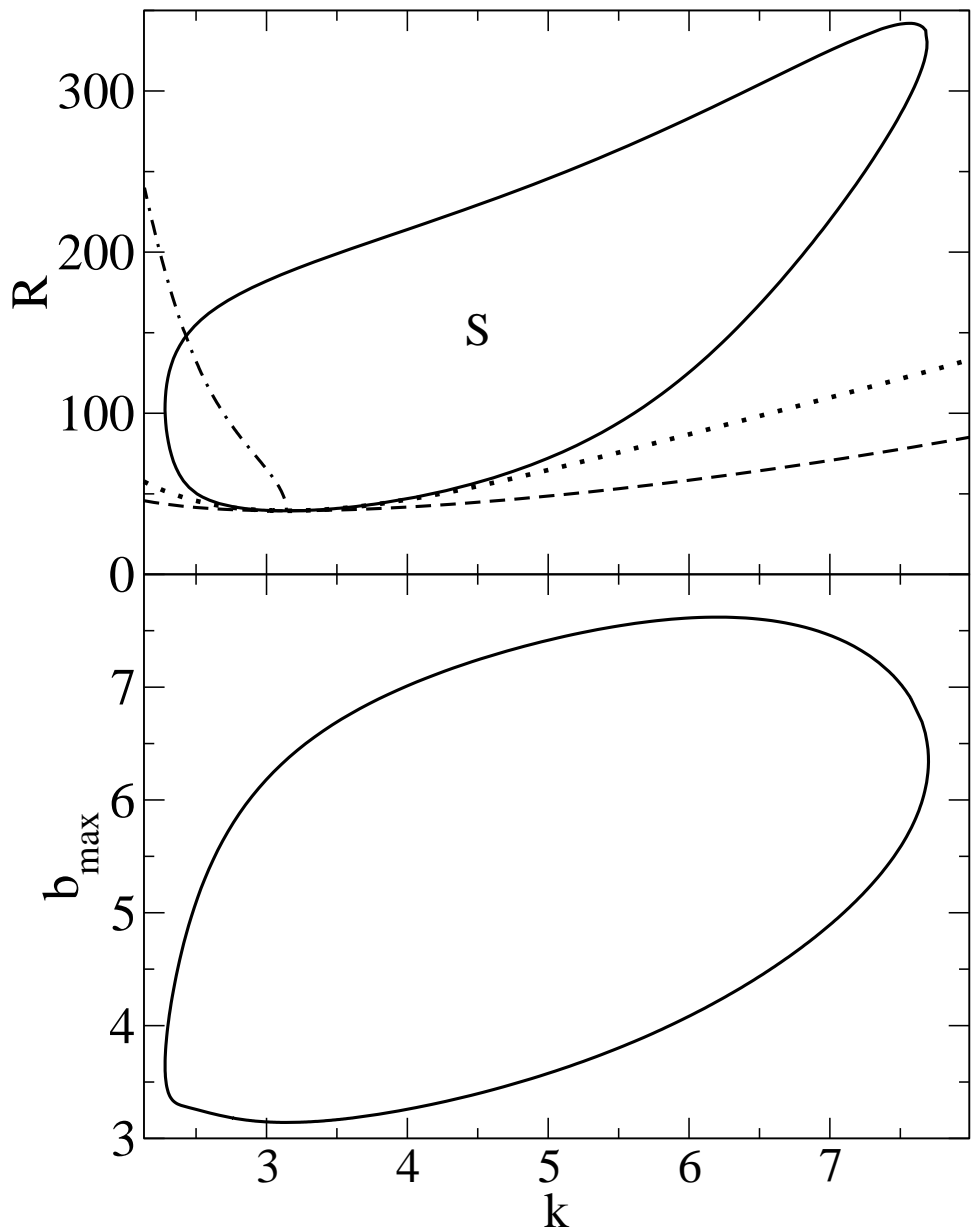}
\caption{Upper part: Stability boundaries in the $(k,R)$-plane for the pure fluid. The Eckhaus boundary is denoted by the dotted curve, the zigzag boundary by the dash-dotted one, the crossroll boundary by the solid one and the neutral curve by the dashed one. The S marks the region of stable rolls. Lower part: $b_{max}$ on the crossroll boundary.}
\label{stab_psi=0} 
\end{figure}

According to Figure \ref{stab_psi=0}, the region of stable rolls is mainly restricted by the crossroll (CR) boundary given by the solid line. Thus, rolls with either too large or too small wavenumbers are destabilised by the CR instability, which causes the growth of rolls perpendicular to the existing pattern. Above $R\approx342=:R_{h}^{0}$, the CR mechanism destabilises the rolls independent of their wavenumber. The CR perturbations are even in $x$-direction and belong to the subclass of $\overline{G}$-perturbations. They have the same periodicity in the $x$-direction as the original rolls so that we find the CR instability for $d=0$. By contrast, the parameter $b$, which represents the wavenumber of the CR perturbation perpendicular to the roll pattern, cannot be fixed to test wether rolls are CR-stable. Instead, to decide this, one has to find the value $b_{max}$ where the most critical eigenvalue reaches its maximum by applying an interpolation procedure. In Figure \ref{stab_psi=0}, the value of $b_{max}$ on the CR boundary is plotted as function of $k$, whereby the upper section of $b_{max}(k)$ belongs to the part of the CR boundary at higher $R$. Near the onset, $b_{max}$ is close to $k_{c}^{0}$. For larger $R$, $b_{max}$ is generally larger since the whole stability region shifts to larger $k$. 
 
The above mentioned value $R_{h}^{0}\approx342$ differs from the one obtained by Straus who states $R_{h}^{0}\approx  380$. Furthermore, our CR boundary is again more restrictive on the small-$k$ side: According to our calculations, rolls with a wavenumber $k< 2.28$ are CR-unstable whereas Straus finds CR-stable rolls with $k\approx1.8$. However, these disagreements decrease when we lower our truncation parameter $N$ and thus we conclude that our results are more precise than those obtained by Straus.

\subsubsection{Eckhaus instability}
The Eckhaus (EC) boundary, denoted by the dotted line in Figure \ref{stab_psi=0}, lies below the CR boundary for all wavenumbers, i.e., the EC perturbations can only grow where the rolls are already CR-unstable. The EC instability tends to establish rolls with a better wavenumber in the direction of the wavevector of the original roll state. The corresponding perturbations are even in x-direction and fall into the subclass of $G$-perturbations. As purely two-dimensional perturbations, they are found at $b=0$. Conveniently, the question of EC stability can be answered by investigating a single point on the $d$-axis near $d=0$  (see \cite{Torre}). Our result for the EC boundary agrees well with the one obtained by Ju\'arez. 
 
Since we have found no other instability mechanisms inside the ZZ- and CR-stable region, we conclude that within this stability region rolls are stable against arbitrary infinitesimal perturbations. In particular, rolls are stable at the critical point $(R_{c}^{0},k_{c}^{0})$. Note that the stability region discussed above is similar to the stability ballon of the clear fluid at large Prandtl numbers for No-slip boundary conditions computed by \cite{BusseClever} as well as for Free-slip boundary conditions computed by \cite{Bolton85}: In each of the three cases, the region of stable rolls is restricted by the ZZ boundary on the small-$k$ side and otherwise by the CR boundary. The similarity of the stability balloons can be understood using a heuristic argument. The influence of the advective term and the time derivative term in the Navier--Stokes equation becomes small for large Prandtl numbers so that it resembles the Darcy equation for small $\gamma_{a}$ except for the fact that the relaxation term in $(\ref{bal}b)$ is replaced by a diffusion term. 

\subsection{Stability boundaries for binary mixtures}

\begin{figure}
\includegraphics[width=0.95\textwidth]{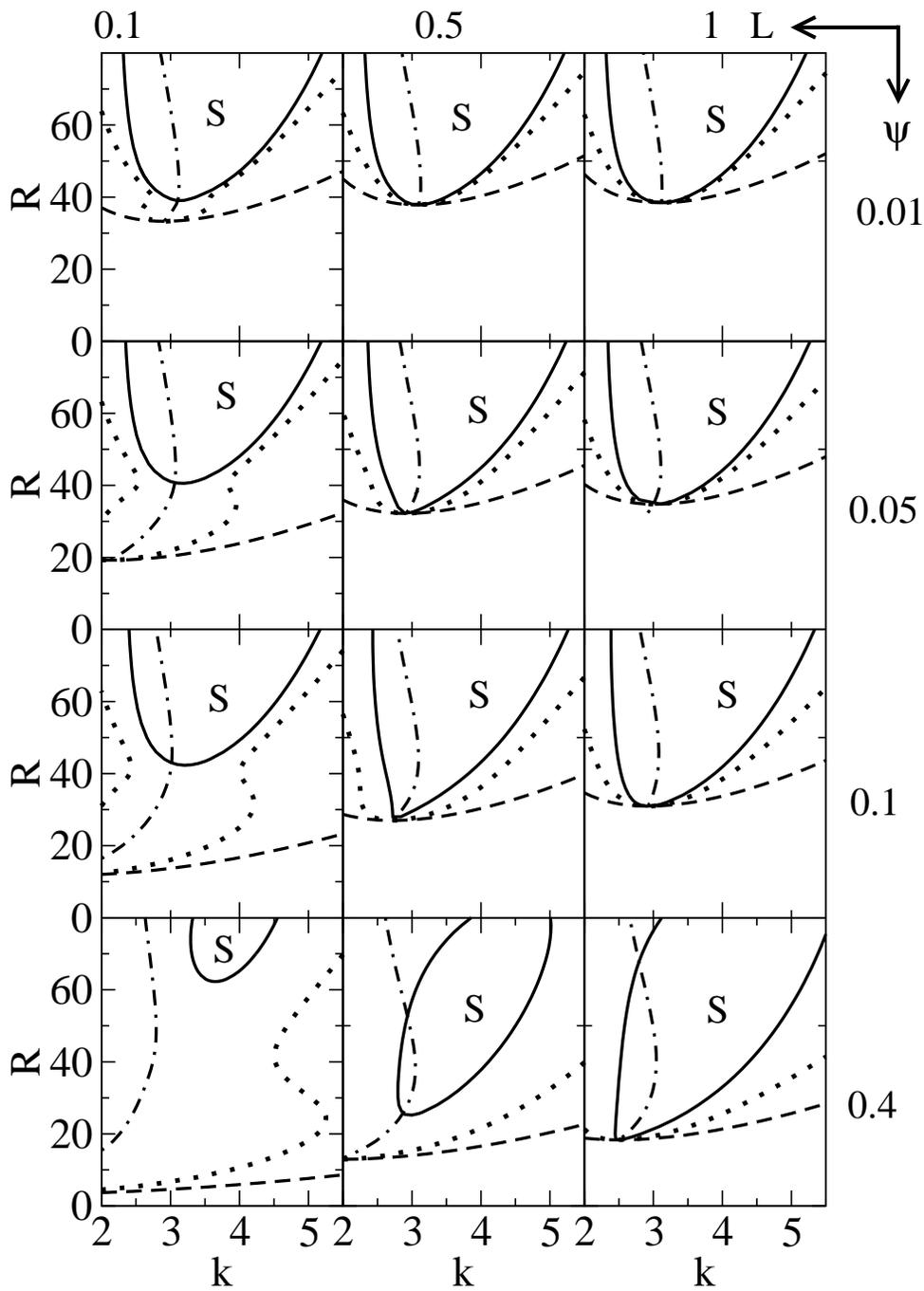}
\caption{Stability boundaries in the $(k,R)$-plane for several combinations of $L$ and $\psi$ below $R=80$. The EC boundary is denoted by the dotted curve, the ZZ boundary by the dash-dotted one, the CR boundary by the solid one and the neutral curve by the dashed one. The S marks the region of stable rolls.}
\label{alle_stab_u} 
\end{figure}

Figure \ref{alle_stab_u} shows the EC, ZZ and CR boundary for binary mixtures below $R=80$. We cover a range of separation ratios ranging from $0.01$ to $0.4$ and Lewis numbers from $0.1$ to $1$. Again, the stability boundaries are found to be in qualitative agreement with those of the clear fluid for large Prandtl numbers, computed by \cite{Huke1}: In the Rayleigh regime, the ZZ boundary $R_{ZZ}(k)$ lies close to the corresponding line of the pure fluid and shifts slightly to smaller $k$ if $\psi$ is increased. At the transition between the Rayleigh and the Soret region, $R_{ZZ}(k)$ bends towards small $k$ (see for instance $L=0.5$ and $\psi=0.05$). By decreasing $R$ further on, the ZZ boundary terminates with a finite slope at the critical point $(R_{c},k_{c})$.
The most important fact regarding the EC boundary is that it lies beyond the CR boundary for all investigated $(L,\psi)$-combinations. Therefore, the EC instability occurs only where the rolls are already unstable to CR perturbations. For $L=1$, which is a typical value for gas mixtures, the CR boundary is still connected to the critical point $(R_{c},k_{c})$. However, for smaller $L$ and sufficiently large $\psi$ the CR boundary detaches from the neutral curve. In this case rolls are no longer a stable form of convection at the onset. We observed square convection at the onset instead as it has also been seen in the clear fluid case by \cite{Lerma1}, \cite{Huke1}, \cite{HW88} and \cite{MS91}.

\begin{figure}
\includegraphics[width=0.95\textwidth]{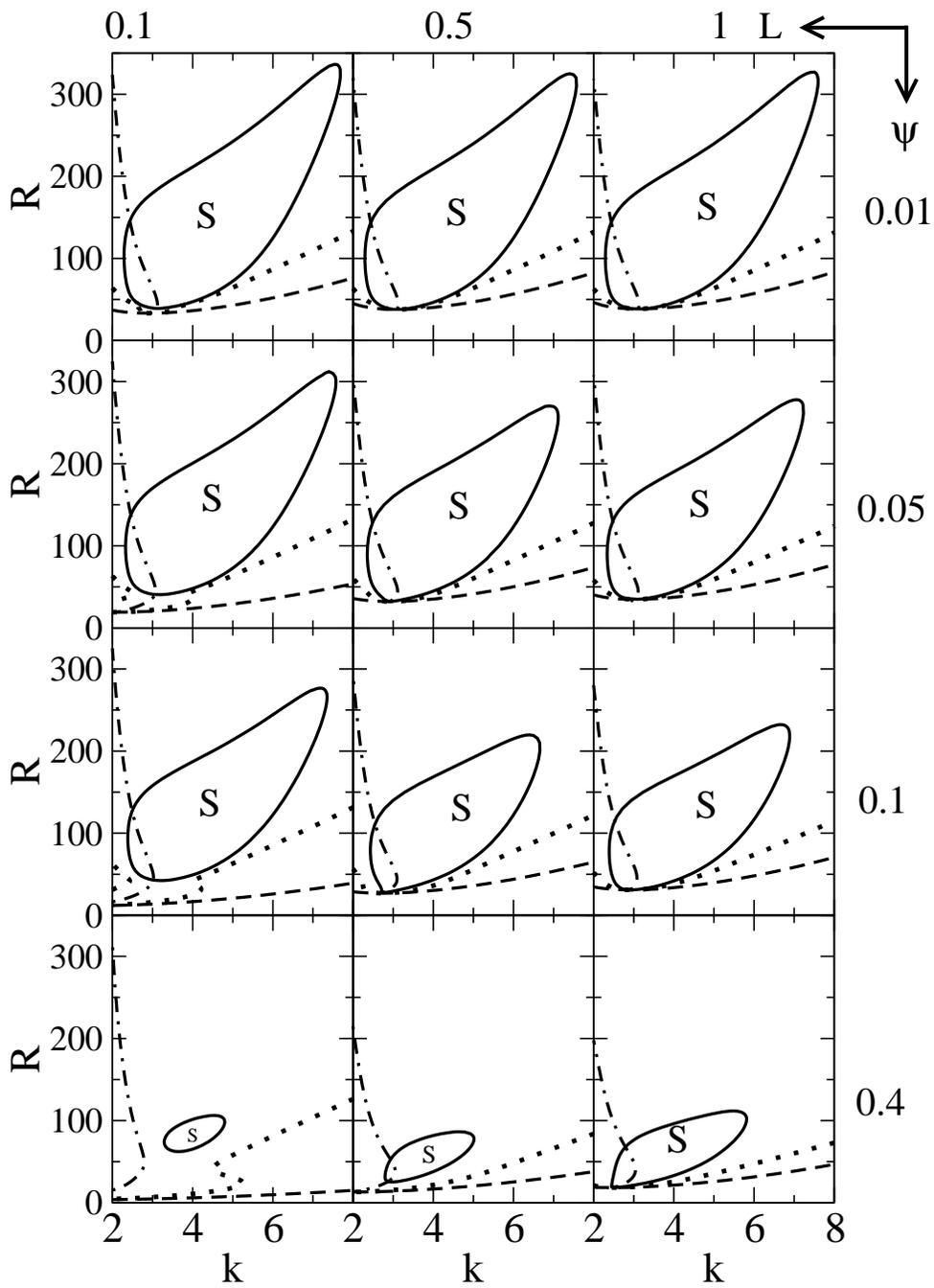}
\caption{Full stability balloons in the $(k,R)$-plane for several combinations of $L$ and $\psi$. The curve style is the same as in Figure \ref{alle_stab_u}. }
\label{alle_stab} 
\end{figure}

The whole $(R,k)$-regions of stable rolls are shown in Figure \ref{alle_stab}. Inside the closed curve defined by the CR boundary rolls are stable to CR perturbations. For large $\psi$ and small $L$, the CR boundary shifts to larger wavenumbers and is the only stability boundary limiting the region of stable rolls (see $L=0.1$ and $\psi=0.4$). Otherwise, the CR-stable region is intersected by the ZZ boundary on the small-$k$ side. For the presented parameter range, we have detected no other instability mechanisms inside the ZZ- and CR-stable region. Thus, the regime of stable rolls is limited exclusively by the ZZ and the CR boundary in analogy to the pure fluid. 

\begin{figure}
\includegraphics[width=0.9\textwidth]{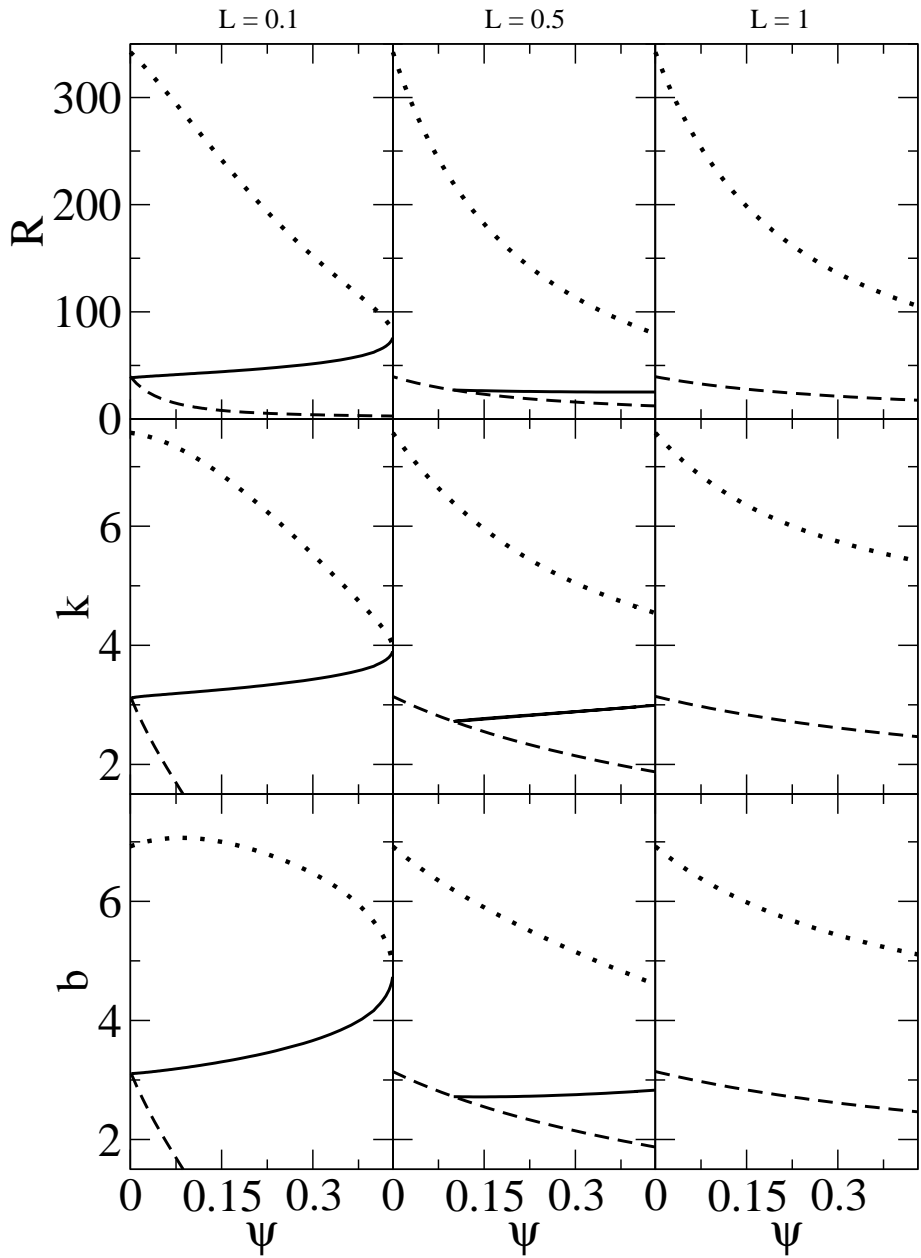}
\caption{First row: For three different Lewis numbers, $R_{h}$, $R_{l}$ and $R_{c}$ denoted by the dotted, solid, dashed lines, respectively, are plotted versus $\psi$. 
         Second row: The corresponding wavenumbers k denoted in analogy to the first row. 
	 Third row: $\psi$ versus  $b_{h}$, $b_{l}$ and $k_{c}$ denoted in analogy to the first row. For further details see text. }
\label{CR} 
\end{figure}

The CR boundary defines the lowest Rayleigh--Darcy number $R_{l}$ as well as the highest one $R_{h}$, for which rolls are stable. As seen from Figure $\ref{CR}$, $R_{h}$ decreases when the separation ratio grows, whereas $R_{l}$ decreases with $\psi$ as soon as the CR boundary becomes detached from the neutral curve. Thus, the region of stable rolls shrinks with increasing Soret effect. Note that this is also valid where the CR boundary is not detached from the ground state  and $R_{l}$ is given by $R_{c}$. Considering the curvature of $R_{l}(\psi)$ and $R_{h}(\psi)$ it seems that CR-stable rolls cannot exist for too large separation ratios. This conjecture is also supported by further numerical tests: We have found no CR-stable rolls for the parameters $L=0.1$ and $\psi=0.5$ on an equidistant grid in the $(R,k)$-plane with $\Delta R =5$ and $\Delta k =0.05$.  

The wavenumber $k_{l}$ ($k_{h}$) of the marginally stable rolls at $R_{l}$ ($R_{h}$) is shown in the second row of Figure \ref{CR} as solid (dotted) lines. Note that $k_{l}$ is equal to $k_{c}$  when the CR boundary is connected to the neutral curve. By increasing $\psi$, $k_{h}$ decreases whereas $k_{l}$ grows when the CR boundary is detached from the neutral curve. The same trend holds for the wavenumber $b_{l}$ ($b_{h}$) in $y$-direction of the most critical CR perturbation at $(R_{l},k_{l})$ ($(R_{h},k_{h})$), which are displayed as solid (dotted) lines in the third row of Figure \ref{CR}. While $b_{l}$ lies near to $k_{l}$ and $b_{h}$ is similar to $k_{h}$, they are not exactly the same.

\subsection{Stability boundaries for binary mixtures with small $L$}

\begin{figure}
\includegraphics[width=0.95\textwidth]{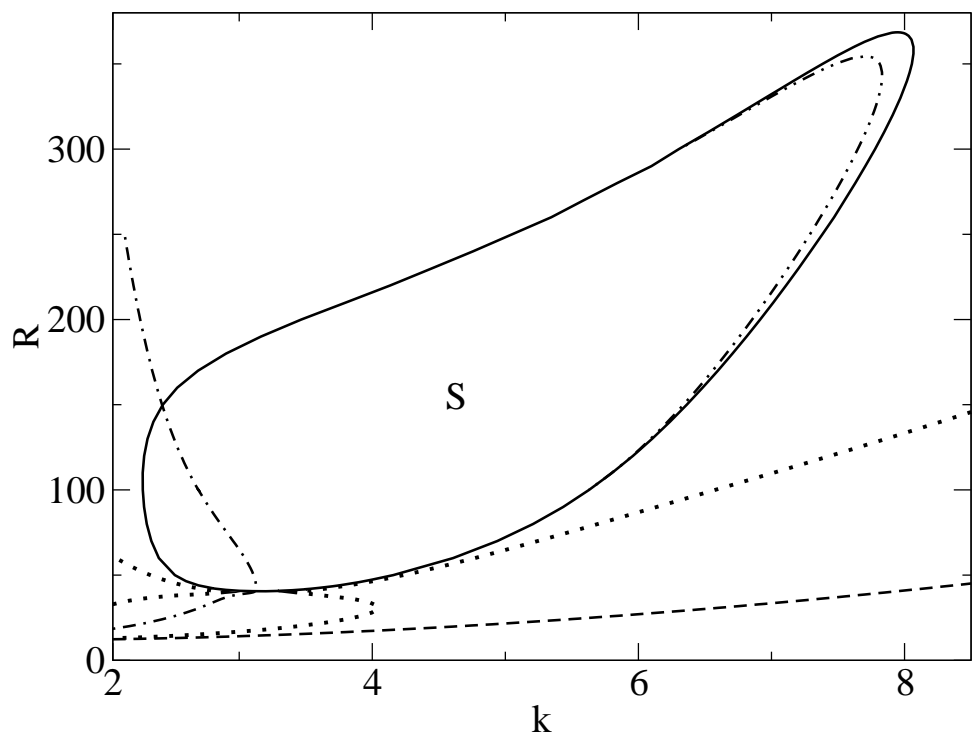}
\caption{Full stability balloons in the $(k,R)$-plane for $L=0.01$ and $\psi=0.01$. The curve style is the same as in Figure \ref{alle_stab_u}, except that the new OCR boundary is denoted by the dashed-double-dotted line.}
\label{small_L}
\end{figure}

Next, we discuss the choice $L=0.01$, which is a typical value for many kinds of liquids. Exemplarily for $\psi=0.01$, the stability balloon is shown in  Figure \ref{small_L}. Whereas the EC, CR and ZZ boundaries remain qualitatively the same as for larger $L$, a new stability boundary marked by a dash-double-dotted line preceeds the CR boundary for sufficiently large $R$. The corresponding eigenvalue reaches its maximum for $d=0$ but for different $b$ in each point of the stability boundary. Moreover, the corresponding perturbations fall into the subclass of $\overline{G}$-perturbations and are even in $x$-direction. As we can see, this new boundary is rather similar to the CR boundary, and it ends on the latter at smaller values of $R$. However, the difference between the CR boundary and the new boundary is that the corresponding eigenvalue of the latter is complex, i.e. the boundary corresponds to an oscillatory perturbation.

\begin{figure}
\includegraphics[width=0.95\textwidth]{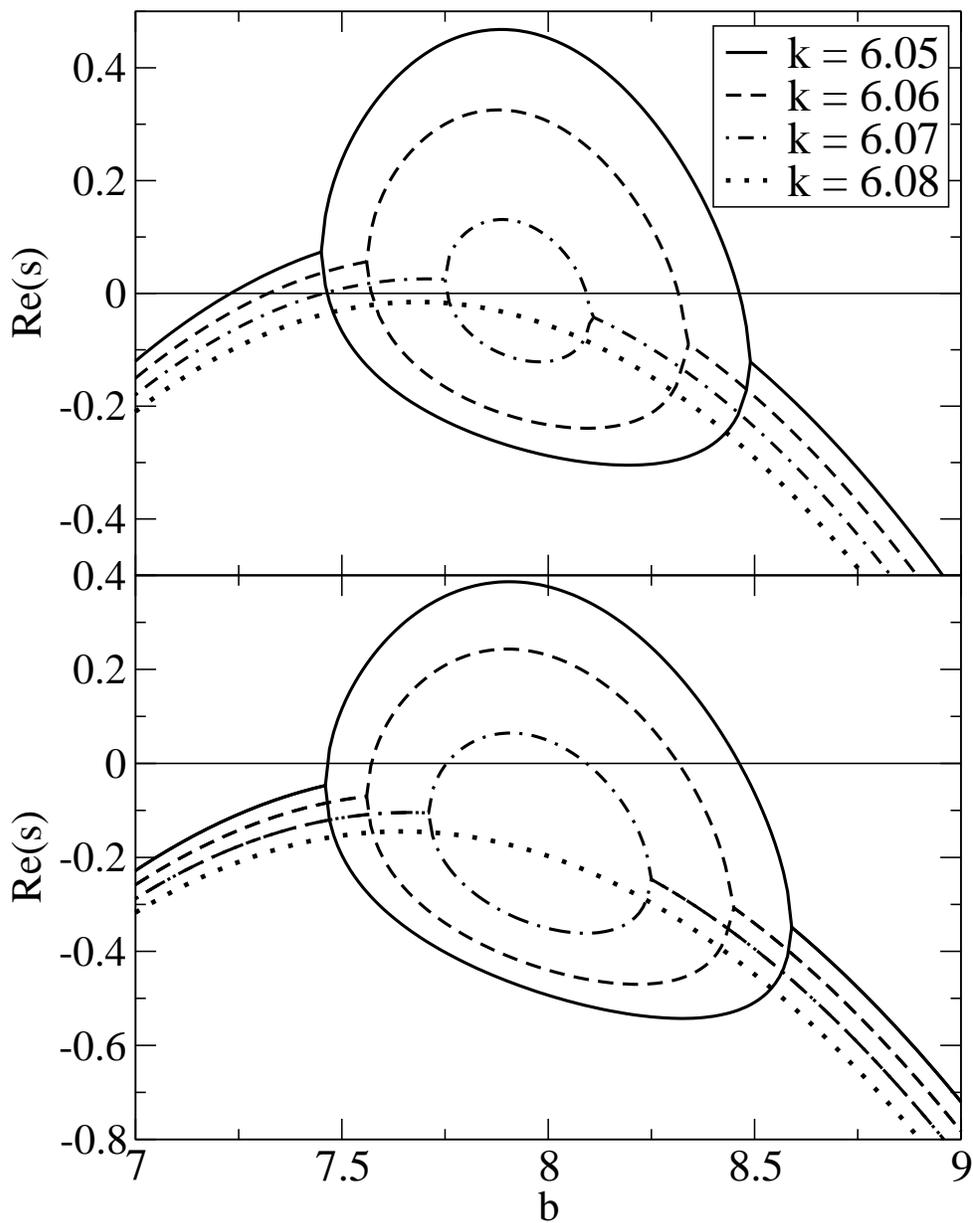}
\caption{Eigenvalues of the OCR and CR boundary at $L=0.01$, $\psi=0.01$, $R=290$, $\phi^{*}=1$ or $\phi^{*}=0.7$ for different values of the wavenumber $k$ as indicated. Plots show the real parts of the two eigenvalues with the greatest real parts.}
\label{EV_OCR}
\end{figure}

In order to clarify the relationship between this new boundary and the CR boundary, Figure \ref{EV_OCR} shows the real parts of the two eigenvalues with the greatest real parts vs. $b$ for different wavenumbers $k$ at parameters $L=0.01$, $\psi=0.01$, $R=290$, $\phi^{*}=1$ and $\phi^{*}=0.7$. For $\phi^{*}=1$, both eigenvalues become positive as a complex pair and thus the corresponding stability boundary is oscillatory. But if $\phi^{*}=0.7$, the eigenvalues split up while they are still negative and only the larger one of the two now real eigenvalues becomes positive. Thus, in this case, the corresponding stability boundary is stationary. Both the new boundary and the CR boundary should thus be understood as one boundary. We call the oscillatory part oscillatory crossroll (OCR) boundary.

As a time dependent phenomenon, the OCR boundary can depend on $\phi^{*}$ and $\gamma_{a}$. With decreasing $\phi^{*}$, the OCR boundary moves towards the CR boundary and the $R$--values increase at which the boundary changes from stationary to oscillatory until the boundary is stationary at all $R$ for sufficiently small $\phi^{*}$. When $\gamma_{a}$ is varied between $10^{-4}$ and $0$, the stability boundaries change negligibly. For example, at fixed $R$, the wavenumbers change less than  $0.1\%$.  Moreover, at fixed $(R,k)$, the real parts of the  eigenvalues grow less than 1\%  with $\gamma_{a}$. The minor influence of $\gamma_{a}$ can be explained as follows: In nondimensional units, the diffusion time of the temperature and the concentration field  are 1 and $1/L=100$, whereas the time scale $\gamma_{a}$ of the velocity field is at least by a factor $10^{4}$ smaller and thus can be neglected.  The case that $\gamma_{a}$ has the same order of magnitude as the diffusion time of the temperature or the concentration field is rather unusual and lies beyond the scope of this work.

Further research turned out to be difficult. For larger values of $\psi$ the boundaries did not converge when increasing the number of modes in our model up to $N=40$, the largest model we used. For large values of $\phi^{*}$ we find the OCR boundary in a large $R$-range while for small values of $\phi^{*}$ the CR boundary seems not to be a closed curve any longer. 

\section{Conclusion}
We studied theoretically roll convection of pure fluids and binary mixtures in the Rayleigh--B\'enard system with porous medium using a Galerkin method. The Soret effect was taken into account whereas we restricted the investigations to mixtures of positive separation ratio.
 
The state of convection has been investigated qualitatively in terms of the streamlines as well as the structure of the temperature and concentration field and quantitatively in terms of the Nusselt number and the mixing number. Whereas the behaviour of the Nusselt and mixing number as well as the behaviour of the temperature and concentration field resembles the situation in the clear fluid case, the behaviour of the streamlines was found to be different. In the Soret regime, the streamlines were deformed since the concentration field transferred its anharmonicity via the buoyancy term into the velocity field. For the clear fluid, this deformation is smoothed by the $\nabla^{2}$-operator in the diffusion term. However, for a fluid in a porous medium, the diffusion term is replaced by a relaxation term in the momentum balance equation deduced from Darcy's law such that the deformation persists. At the transition to the Rayleigh regime, the deformation diminishes since form there on the buoyancy force is generated rather by the harmonic temperature field than by the anharmonic concentration field. 

Also, we investigated the stability of rolls against arbitrary infinitesimal perturbations. The stability region for a pure fluid was found to be restricted by the ZZ instability and the CR instability only which is in qualitative agreement with the one of \cite{Straus}. However, compared to \cite{Straus}, our stability balloon is more restrictive on the small-$k$ side and the highest $R$ for which rolls should be stable against infinitesimal perturbation is calculated to about $342$ instead of $380$ stated by Straus. 

In a next step, we extended the stability analysis to binary mixtures with $\psi>0$. For not too small Lewis numbers ($L\geq0.1$), the ZZ and the CR instability mechanism were found again to be the only relevant mechanisms. The corresponding stability boundaries changed similarly to the clear fluid case: In the Rayleigh regime, the ZZ boundary was shifted slightly to smaller $k$ when we increased $\psi$ and bent towards smaller $k$ at the transition to the Soret regime. The CR boundary was found to be detached from the neutral curve for sufficiently large $\psi$. Moreover, the CR stable region shrinked for growing separation ratio. Thus, for example, we found no stable rolls for $\psi\geq0.5$ and $L=0.1$. 

For small Lewis numbers such as $L=0.01$, we observed a change in the behaviour of the CR boundary. For strong thermal driving, the critical eigenvalue corresponding to the CR boundary becomes complex. Therefore, the boundary changes to an oscillatory one. Up to now, this phenomenon has not been observed in the clear fluid system. The point at which the boundary changes from stationary to oscillatory is dependent of $\psi$ and $\phi^{*}$. When $\phi^{*}$ becomes small enough, the CR boundary is stationary for all values of $R$. Further research for larger $\psi$ turned out to be difficult because the computed boundaries did not converge. For $\psi=0.1$ we found that the CR boundary becomes oscillatory in a larger $R$-range and that the stationary CR boundary does not seem to form a closed curve any longer.

Motivated by the work of Vadasz (for example \cite{Vad99}), we retained the time derivative in the momentum balance provided with the correction factor $\gamma_{a}$. Whereas, this term was essential in the stability analyses of Vadasz, we did not observe any noteworthy influence on our results as $\gamma_{a}$ was varied  in the range from 0 to $10^{-4}$.
\newpage

\bibliographystyle{jfm}
\bibliography{rollen}

\begin{thebibliography}{39}
\expandafter\ifx\csname natexlab\endcsname\relax\def\natexlab#1{#1}\fi

\bibitem[Ahlers \& Rehberg(1986)]{Ahlers2}
{\sc Ahlers, G. \& Rehberg, I.} 1986 {C}onvection in a {B}inary {M}ixture
  {H}eated from {B}elow. {\em Phys.~Rev.~Lett.\/} {\bf 56}~(13), 1373--1376.

\bibitem[Barten {\em et~al.\/}(1989)Barten, L{\"u}cke, Hort \& Kamps]{Barten2}
{\sc Barten, W., L{\"u}cke, M., Hort, W. \& Kamps, M.} 1989 {F}ully developed
  traveling-wave convection in binary fluid mixtures. {\em Phys.~Rev.~Lett.\/}
  {\bf 63}~(4), 376--379.

\bibitem[Barten {\em et~al.\/}(1995)Barten, L{\"u}cke, Kamps \&
  Schmitz]{Barten1}
{\sc Barten, W., L{\"u}cke, M., Kamps, M. \& Schmitz, R.} 1995 {C}onvection in
  binary fluid mixtures. {I}. {E}xtended traveling-wave and stationary states.
  {\em Phys.~Rev.~E\/} {\bf 51}~(6), 5636--5661.

\bibitem[Bolton \& Busse(1985)]{Bolton85}
{\sc Bolton, E.~W. \& Busse, F.~H.} 1985 {S}tability of convection rolls in a
  layer with stress-free boundaries. {\em J.~Fluid~Mech.\/} {\bf 150},
  487--498.

\bibitem[Bolton {\em et~al.\/}(1986)Bolton, Busse \& Clever]{Bo86}
{\sc Bolton, E.~W., Busse, F.~H. \& Clever, R.~M.} 1986 {O}scillatory
  instabilities of convection rolls at intermediate {P}randtl numbers. {\em
  J.~Fluid~Mech.\/} {\bf 164}, 469--485.

\bibitem[Busse \& Clever(1979)]{BusseClever}
{\sc Busse, F.~H. \& Clever, R.~M.} 1979 {I}nstabilities of convection rolls in
  a fluid of moderate {P}randtl number. {\em J.~Fluid~Mech.\/} {\bf 91},
  319--335.

\bibitem[Busse \& Whitehead(1971)]{Busse71}
{\sc Busse, F.~H. \& Whitehead, J.~A.} 1971 {I}nstabilities of convection rolls
  in a high {P}randtl number fluid. {\em J.~Fluid~Mech.\/} {\bf 47}, 305--320.

\bibitem[Charrier-Mojtabi {\em et~al.\/}(2007)Charrier-Mojtabi, Elhajjar \&
  Mojtabi]{Mo07}
{\sc Charrier-Mojtabi, M.-C., Elhajjar, B. \& Mojtabi, A.} 2007 {A}nalytical
  and numerical stability analysis of {S}oret--driven convection in a
  horizontal porous layer. {\em Physics~of~Fluids\/} {\bf 19}~(12), 124104.

\bibitem[Clever \& Busse(1990)]{ClBu90}
{\sc Clever, R.~M. \& Busse, F.~H.} 1990 {C}onvection at very low {P}randtl
  numbers. {\em Physics~of~Fluids\/} {\bf 2}~(3), 334--339.

\bibitem[Croquette \& Williams(1989)]{Croquette89a}
{\sc Croquette, V. \& Williams, H.} 1989 {N}onlinear competition between waves
  on convective rolls. {\em Phys.~Rev.~A\/} {\bf 39}~(5), 2765--2768.

\bibitem[Cross \& Kim(1988)]{CrossKim}
{\sc Cross, M.~C. \& Kim, K.} 1988 {L}inear instability and the codimension-2
  region in binary fluid convection between rigid impermeable boundaries. {\em
  Phys.~Rev.~A\/} {\bf 37}~(10), 3909--3920.

\bibitem[De~La Torre~Ju{\'a}rez \& Busse(1995)]{Torre}
{\sc De~La Torre~Ju{\'a}rez, M. \& Busse, F.~H.} 1995 {S}tability of
  two-dimensional convection in a fluid-saturated porous medium. {\em
  J.~Fluid~Mech.\/} {\bf 292}, 305--323.

\bibitem[Dominguez-Lerma {\em et~al.\/}(1995)Dominguez-Lerma, Ahlers \&
  Cannell]{Lerma1}
{\sc Dominguez-Lerma, M.~A., Ahlers, G. \& Cannell, D.~S.} 1995
  {R}ayleigh--{B}{\'e}nard convection in binary mixtures with separation ratios
  near zero. {\em Phys.~Rev.~E\/} {\bf 52}~(6), 6159--6174.

\bibitem[Eaton {\em et~al.\/}(1991)Eaton, Ohlsen, Yamamoto, Surko, Barten,
  L{\"u}cke, Kamps \& Kolodner]{Eaton}
{\sc Eaton, K.~D., Ohlsen, D.~R., Yamamoto, S.~Y., Surko, C.~M., Barten, W.,
  L{\"u}cke, M., Kamps, M. \& Kolodner, P.} 1991 {C}oncentration field in
  traveling-wave and stationary convection in fluid mixtures. {\em
  Phys.~Rev.~A\/} {\bf 43}~(12), 7105--7108.

\bibitem[Elhajjar {\em et~al.\/}(2008)Elhajjar, Charrier-Mojtabi \&
  Mojtabi]{Mo08}
{\sc Elhajjar, B., Charrier-Mojtabi, M.-C. \& Mojtabi, M.} 2008 {S}eparation of
  a binary fluid mixture in a porous horizontal cavity. {\em Phys.~Rev.~E\/}
  {\bf 77}~(2), 026310.

\bibitem[F{\"u}tterer \& L{\"u}cke(2002)]{Fuetterer1}
{\sc F{\"u}tterer, C. \& L{\"u}cke, M.} 2002 {G}rowth of binary fluid
  convection: {R}ole of the concentration field. {\em Phys.~Rev.~E\/} {\bf
  65}~(3), 036315.

\bibitem[Hollinger(1996)]{DissHollinger}
{\sc Hollinger, St.} 1996 {T}heorie der ausgedehnten station{\"a}ren und
  wandernden {K}onvektion in bin{\"a}ren {F}luidmischungen. PhD thesis,
  Universit{\"a}t des Saarlandes, unpublished.

\bibitem[Howle {\em et~al.\/}(1997)Howle, Behringer \& Georgiadis]{How97}
{\sc Howle, L.~E., Behringer, R.~P. \& Georgiadis, J.~G.} 1997 {C}onvection and
  flow in porous media. {P}art 2. {V}isualization by shadowgraph. {\em
  J.~Fluid~Mech.\/} {\bf 332}, 247--262.

\bibitem[Huke. \& L{\"u}cke(2002)]{Huke2}
{\sc Huke., B. \& L{\"u}cke, M.} 2002 {C}onvective {P}atterns in {B}inary
  {F}luid {M}ixtures with {P}ositive {S}eparation {R}atios. In {\em {T}hermal
  {N}onequilibrium {P}henomena in {F}luid {M}ixtures\/} (ed. W.~K{\"o}hler \&
  S.~Wiegand), {\em {L}ecture {N}otes in {P}hysics {M}onographs\/}, vol. 584,
  pp. 334--354. Berlin: Springer-Verlag.

\bibitem[Huke {\em et~al.\/}(2000)Huke, L{\"u}cke, B{\"u}chel \& Jung]{Huke1}
{\sc Huke, B., L{\"u}cke, M., B{\"u}chel, P. \& Jung, Ch.} 2000 {S}tability
  boundaries of roll and square convection in binary fluid mixtures with
  positive separation ratio. {\em J.~Fluid~Mech.\/} {\bf 408}, 121--147.

\bibitem[Knobloch \& Moore(1988)]{KnoblochMoore}
{\sc Knobloch, E. \& Moore, D.~R.} 1988 {L}inear stability of experimental
  {S}oret convection. {\em Phys.~Rev.~A\/} {\bf 37}~(3), 860--870.

\bibitem[Kolodner {\em et~al.\/}(1986)Kolodner, Passner, Surko \&
  Walden]{KolodnerPRL}
{\sc Kolodner, P., Passner, A., Surko, C.~M. \& Walden, R.~W.} 1986 {O}nset of
  {O}scillatory {C}onvection in a {B}inary {F}luid {M}ixture. {\em
  Phys.~Rev.~Lett.\/} {\bf 56}, 2621--2624.

\bibitem[Le~Gal {\em et~al.\/}(1985)Le~Gal, Pocheau \& Croquette]{Gal}
{\sc Le~Gal, P., Pocheau, A. \& Croquette, V.} 1985 {S}quare versus {R}oll
  {P}attern at {C}onvective {T}hreshold. {\em Phys.~Rev.~Lett.\/} {\bf
  54}~(23), 2501--2504.

\bibitem[Liu \& Ahlers(1997)]{Liu}
{\sc Liu, J. \& Ahlers, G.} 1997 {R}ayleigh--{B}{\'e}nard convection in
  binary-gas mixtures: {T}hermophysical properties and the onset of convection.
  {\em Phys.~Rev.~E\/} {\bf 55}~(6), 6950--6968.

\bibitem[Moses \& Steinberg(1991)]{MS91}
{\sc Moses, E. \& Steinberg, V.} 1991 {S}tationary convection in a binary
  mixture. {\em Phys.~Rev.~A\/} {\bf 43}~(2), 707--722.

\bibitem[M{\"u}ller \& L{\"u}cke(1988)]{HW88}
{\sc M{\"u}ller, H.~W. \& L{\"u}cke, M.} 1988 {C}ompetition between roll and
  square convection patterns in binary mixtures. {\em Phys.~Rev.~A\/} {\bf
  38}~(6), 2965--2974.

\bibitem[Nield \& Bejan(2006)]{Ni06}
{\sc Nield, D.~A. \& Bejan, A.}, ed. 2006 {\em {C}onvection in {P}orous
  {M}edia\/}. New York: Springer-Verlag.

\bibitem[Platten(2006)]{Pl06}
{\sc Platten, J.~K.} 2006 {T}he {S}oret--{E}ffect: {A} {R}eview of {R}ecent
  {E}xperimental {R}esults. {\em J.~of~Applied~Mech.\/} {\bf 73}~(1), 5--15.

\bibitem[Schl{\"u}ter {\em et~al.\/}(1965)Schl{\"u}ter, Lortz \&
  Busse]{Schlueter65}
{\sc Schl{\"u}ter, A., Lortz, D. \& Busse, F.~H.} 1965 {O}n the stability of
  steady finite amplitude convection. {\em J.~Fluid~Mech.\/} {\bf 23},
  129--144.

\bibitem[Sch{\"o}pf \& Zimmermann(1993)]{Schoepf}
{\sc Sch{\"o}pf, W. \& Zimmermann, W.} 1993 {C}onvection in binary fluids:
  {A}mplitude equations, codimension-2 bifurcation, and thermal fluctuations.
  {\em Phys.~Rev.~E\/} {\bf 47}~(3), 1739--1764.

\bibitem[Shattuck {\em et~al.\/}(1997)Shattuck, Behringer, Johnson \&
  Georgiadis]{Sha97}
{\sc Shattuck, M.~D., Behringer, R.~P., Johnson, G.~A. \& Georgiadis, J.~G.}
  1997 {C}onvection and flow in porous media. {P}art 1. {V}isualization by
  magnetic resonance imaging. {\em J.~Fluid~Mech.\/} {\bf 332}, 215--245.

\bibitem[Sovran {\em et~al.\/}(2001)Sovran, Charrier-Mojtabi \& Mojtabi]{So01}
{\sc Sovran, O., Charrier-Mojtabi, M.-C. \& Mojtabi, M.} 2001 {N}aissance de la
  convection thermo--solutale en couche poreuse infinie avec effet {S}oret.
  {\em Comptes rendus de l'{A}cad{\'e}mie des sciences. {S}{\'e}rie {II}b,
  {M}{\'e}canique\/} {\bf 329}~(4), 287--293.

\bibitem[Straus(1974)]{Straus}
{\sc Straus, J.~M.} 1974 {L}arge amplitude convection in porous media. {\em
  J.~Fluid~Mech.\/} {\bf 64}, 51--63.

\bibitem[Touiri {\em et~al.\/}(1996)Touiri, Platten \& Chavepeyer]{Touiri}
{\sc Touiri, H., Platten, J.~K. \& Chavepeyer, G.} 1996 {E}ffect of the
  separation ratio on the transition between travelling waves and steady
  convection in the two--component {R}ayleigh--{B}{\'e}nard problem. {\em
  Eur.~J.~Mech.~B\/} {\bf 15}~(2), 241--257.

\bibitem[Vadasz \& Olek(1999)]{Vad99}
{\sc Vadasz, P. \& Olek, S.} 1999 {W}eak {T}urbulence and {C}haos for {L}ow
  {P}randtl {N}umber {G}ravity {D}riven {C}onvection in {P}orous {M}edia. {\em
  Transport~in~Porous~Media\/} {\bf 37}~(1), 69--91.

\bibitem[Vadasz \& Olek(2000)]{Vad00}
{\sc Vadasz, P. \& Olek, S.} 2000 {R}oute to {C}haos for {M}oderate {P}randtl
  {N}umber {C}onvection in a {P}orous {L}ayer {H}eated from {B}elow. {\em
  Transport~in~Porous~Media\/} {\bf 41}, 211--239.

\bibitem[Vafai(2005)]{Vaf05}
{\sc Vafai, K.}, ed. 2005 {\em {H}andbook of {P}orous {M}edia\/}. New York:
  Springer-Verlag.

\bibitem[Veronis(1966)]{Ve66}
{\sc Veronis, G.} 1966 {L}arge--amplitude {B}{\'e}nard convection. {\em
  J.~Fluid~Mech.\/} {\bf 26}, 49--68.

\bibitem[Walden {\em et~al.\/}(1985)Walden, Kolodner, Passner \& Surko]{Walden}
{\sc Walden, R.~W., Kolodner, P., Passner, A. \& Surko, C.~M.} 1985 {T}raveling
  waves and chaos in convection in binary fluid mixtures. {\em
  Phys.~Rev.~Lett.\/} {\bf 55}~(5), 496--499.

\end{thebibliography}

\end{document}